\documentclass[manuscript]{aastex}

\slugcomment{submitted to \apj}

\begin{document}

\title{Structural       Parameters        of      the  M87       Globular
  Clusters\altaffilmark{1}}

%---------------------------------------------------------------------

\author{
Juan P. Madrid\altaffilmark{2}, 
William E. Harris\altaffilmark{2},
John P. Blakeslee \altaffilmark{3},
Mat\'ias G\'omez \altaffilmark{4} 
}

\altaffiltext{1}{Based on  observations made with  the NASA/ESA Hubble
Space Telescope,  obtained at  the Space Telescope  Science Institute,
which is operated  by the Association of Universities  for Research in
Astronomy, Inc., under NASA  contract NAS 5-26555.  These observations
are associated with program 10543.}

\altaffiltext{2}{Department   of  Physics   and   Astronomy,  McMaster
  University, Hamilton ON L8S 4M1, Canada}

\altaffiltext{3}{Herzberg Institute of  Astrophysics, Victoria, BC V9E
  2E7, Canada}

\altaffiltext{4}{Departamento   de  Ciencias   Fisicas,   Facultad  de
  Ingenieria, Universidad Andres Bello, Concepci\'on, Chile}

%---------------------------------------------------------------------

\begin{abstract}

  We derive structural parameters for $\sim 2000$ globular clusters in
  the  giant Virgo elliptical  M87 using  extremely deep  Hubble Space
  Telescope images in F606W (V)  and F814W (I) taken with the ACS/WFC.
  The cluster  scale sizes (half-light radii  $r_h$) and ellipticities
  are  determined from  PSF-convolved King-model  profile  fitting. We
  find that  the $r_h$ distribution closely resembles  the inner Milky
  Way clusters, peaking  at $r_h \simeq 2.5$ pc  and with virtually no
  clusters  more  compact than  $r_h  \simeq  1$  pc.  The  metal-poor
  clusters have  on average an  $r_h$ 24\% larger than  the metal-rich
  ones. The cluster scale size shows a gradual and noticeable increase
  with galactocentric distance.  Clusters  are very slightly larger in
  the bluer waveband $V$, a possible  hint that we may be beginning to
  see the  effects of mass  segregation within the clusters.   We also
  derived  a color  magnitude  diagram for  the  M87 globular  cluster
  system which show a striking bimodal distribution.

\end{abstract}  


%---------------------------------------------------------------------

\section{Introduction} % Section 1

Messier 87 (Virgo A, and  NGC 4486) is the central dominant elliptical
galaxy  of the  Virgo Cluster  located 16.7  Megaparsecs away  from us
(Blakeslee et al.  2009).  M87 is one of the most fascinating and best
studied celestial  objects and due  to its relative close  distance it
has been  a prime target  of observations that have  provided precious
clues for  the understanding of  black hole physics,  accretion disks,
radio galaxies, and jets, among many other fields.

M87  has  also   played  a  crucial  role  in   the  understanding  of
extragalactic  globular  cluster  systems  (GCS).   One  of  the  most
prominent characteristics of this galaxy is its extremely large number
of globular clusters.  Recent ground-based observations found that the
M87 GCS has in total more than 14000 members (Tamura et al. 2006). M87
is  also the  paradigmatic  high specific  frequency ($S_N$)  globular
cluster system  (McLaughlin et al.  1994, Harris  2009).  The specific
frequency represents the number of clusters per unit galaxy luminosity
(Harris \&  van den Bergh  1981) and for  M87 $S_N$=14, a  value three
times  higher than other  Virgo giant  ellipticals (McLaughlin  et al.
1994, Peng et al.  2008).  This large sample of globular clusters (GC)
constitutes a  superb basis for  carrying out many  studies, including
now accurate measurements of their structural parameters.

The  large number  of globular  clusters  belonging to  M87 was  first
noticed  in  1955  (Baum  1955)  but the  study  of  their  structural
parameters, e.  g.  characteristic sizes, ellipticities, luminosities,
is  enabled  only by  the  superior  resolution  of the  Hubble  Space
Telescope  (HST).  A  typical size  expected for  a  globular cluster,
namely $r_{eff}\sim$  2 to  3 parsecs in  radius, corresponds  to only
about  0.035\arcsec  at  the  Virgo  distance, or  one  pixel  of  the
instruments onboard HST  (Larsen et al.  2001). Previous  work of this
type  for the  globular  clusters in  the  Virgo members  was done  by
Jord\'an et al.   (2005, 2009) with Advanced Camera  for Surveys (ACS)
data  but with  much  shorter-exposure  data than  we  use here.  Even
earlier  work for  M87 (Kundu  at al.  1999) used  the WFPC2  with its
larger pixel scale and smaller field of view, and is now superseded by
the capabilities of the ACS.

In this study we use one of the deepest datasets in the HST archive to
derive  the structural  parameters for  globular clusters  in  M87.  A
large, multi-orbit  program imaging the  central region of  the galaxy
was carried out with the ACS and produced exceptionally deep images of
the  inner region  of  M87 (GO  10543,  PI: Baltz),  a region  heavily
illuminated by galaxy light and difficult to image to such detail with
a telescope of  lower resolution or image quality.   In this very deep
dataset  more than  two thousand  sources are  globular  clusters, and
their radial profiles are distinctly and obviously larger than the PSF
of the HST/ACS.   We use the fact that these  clusters are resolved to
measure  their structural  parameters  with {\sc  ishape}, a  software
specially  designed to  derive  these parameters  for barely  resolved
objects (Larsen 1999).

The morphology  of individual globular clusters is  characterized by a
number of standard parameters,  perhaps most importantly the effective
radius $r_h$,  the radius that  contains half the total  luminosity of
the cluster.  The effective  radius remains virtually constant through
up to  ten relaxation times  and thus plays  an important role  in the
understanding of  evolutionary processes of  globular clusters systems
back nearly to the proto-cluster stage (Spitzer \& Thuan 1972, Aarseth
\&  Heggie  1998,  G\'omez  et  al.   2006).   The  M87  clusters  are
associated with the  very dense environment of the  Virgo Cluster core
in a  supergiant elliptical and thus their  structural parameters give
us the chance  to probe a very different environment  than the ones in
the Milky Way.

This paper is  structured as follows: \S 2  describes the observations
and data  reduction, \S 3  describes the modeling  of the PSF  used to
derived  the structural  parameters  presented in  \S  4, in  \S 5  we
discuss the photometry and the  color-magnitude diagram of the M87 GCS
while \S 6 does a comparison  with another recent study using the same
dataset, and in  \S 7 we highlight the most  important results of this
study.

We   use  throughout   this   paper  H$_0$=71   km.s$^{-1}$Mpc$^{-1}$,
$\Omega_M=0.27$, and $\Omega_\Lambda=0.73$. At the distance of M87,
1$\arcsec$ corresponds to 80 pc (Wright 2006).

%---------------------------------------------------------------------

\section{Observations and Data Reduction} % Section 2

Our images were taken with the ACS Wide Field Channel (WFC), which has
a $202\arcsec  \times 202\arcsec$ field of  view and a  pixel scale of
$0.05\arcsec$  (Mack et  al.  2003,  Gonzaga  et al.   2005). The  ACS
observations that we analyze were taken in two filters: F606W (wide V)
and  F814W   (equivalent  to  I).   The  original   program  of  these
observations  was aimed  at finding  microlensing events  towards M87.
These microlensing events are expected  to be red (Baltz et al.  2004)
therefore the  observations were heavily  weighted to F814W,  with 205
images  and a  total of  73800 seconds.   The F606W  observations were
meant to provide mostly  supplementary color information, and comprise
49 images totaling 24500 seconds.  The data was taken over a period of
71  days, from 2005  Dec 24  to 2006  Mar 05  in 61  different visits.
These  observations  were obtained  following  a  dither pattern  that
allows sub-pixel drizzling.

We  retrieved the flatfielded  science files  {\sc flt.fits}  from the
Multimission  Archive at  Space  Telescope (MAST).   These files  were
prepared through the standard  ``on the fly reprocessing'' (OTFR). The
OTFR system processes  the raw HST data using  up-to-date software and
calibration files.  For the  Advanced Camera for Surveys the automatic
reduction  and  calibration  pipeline  ({\sc  calacs})  performs  bias
subtraction,  dark   current  subtraction,  and   flatfielding  before
delivery of each {\sc flt.fits} file (Sirianni et al. 2005).

Subsequent combination  of the images  was done with the  ACS Pipeline
Science Investigation Software (APSIS) and  Pyraf.  We use APSIS to do
distortion correction,  image registration, image  combination, cosmic
ray  rejection, and  drizzling  (Blakeslee et  al.   2003). APSIS  was
developed by the ACS science team and proved to be more effective than
{\sc  multidrizzle},  the alternative  software,  at registering  this
large  data set taken  during different  epochs.  Using  Python, APSIS
performs  subpixel drizzling,  and  we chose  to  construct the  final
science  images with  a pixel  scale of  0.035\arcsec/pixel. We  use a
gaussian  function  as  interpolation  kernel  and  a  pixel  fraction
(pixfrac) of 1.

As a test  to the solidity of our  results, all measurements described
below were  also performed on an  image with a narrower  PSF. An image
with  a 10\%  narrower PSF  was  built by  using a  smaller value  for
pixfrac (0.5  instead of  1.0) during the  data reduction.   The final
cluster  sizes  measured  from  these two  different  reductions  were
identical. Using  a different fitting  code ({\sc kingphot}),  Peng et
al.  (2009) come  to the same conclusion that we  do here, namely that
the differences  for the magnitudes  and sizes are  negligible between
the two reductions.

From the  master combined F606W and  F814W images we  created a median
subtracted image to eliminate the  galaxy light in order to facilitate
the detection of the position of globular clusters on our images.  The
pyraf task  {\sc daofind} did  a fine job detecting  globular clusters
and  ignoring  brightness fluctuations  and  background galaxies.

We performed a careful visual  inspection of the frame to verify these
detections.  We  manually removed spurious sources along  the edges of
the  different  knots  belonging  to  the prominent  M87  jet.   Sharp
gradients in  brightness are often  misinterpreted as sources  by {\sc
  daofind}. We also removed spurious detections along the edges of the
image. We detect  a total of 2010 objects on our  images almost all of
which are  globular clusters.  In  Figure 1 we  show a section  of the
median  subtracted image  with our  detections of  cluster candidates.
Given the  very large  number of  clusters in our  sample a  few field
contaminants have  a negligible impact  on our results (Larsen  et al.
2001).  We  should note that  we performed all  measurements presented
below on the original image, the median subtracted image was used only
to locate  the clusters. Systematics  introduced by a routine  such as
{\sc median} or {\sc ellipse}  are difficult to evaluate and therefore
difficult to correct even though they may be small.

%---------------------------------------------------------------------
%---------------------------------------------------------------------

\section{PSF modeling}

An  accurate  PSF is  crucial  for  our  purpose of  deriving  precise
structural  parameters for the  GCs, given  that we  are working  in a
regime where the  GC intrinsic sizes $r_h$ are similar  to the FWHM of
the  PSF  itself.   In  principle,  PSFs can  be  constructed  through
modeling  techniques such  as TinyTim,  but extensive  discussions of
this  approach in  recent papers  directed  towards the  same goal  of
measuring  GC sizes  in  relatively nearby  galaxies  imaged with  ACS
(Spitler  et  al  2006,  Georgiev  et  al.   2008)  show  considerable
difficulty matching the models to empirical PSFs derived directly from
stars on  the images.  Working  from this experience, we  have adopted
the approach  of using purely empirical  PSFs built from  stars on the
master F814W and F606W images described above.

An  immediate practical  issue is  to  find candidate  stars on  these
images: of the  more than 2000 detected objects  on the field, roughly
98\%  are globular clusters  belonging to  M87 and  these must  all be
weeded out.   Fortunately, the  true size of  the stars  is accurately
known  beforehand:  for  properly   co-added  images,  the  HST  image
resolution is  $\simeq 0.095''$, and  thus we expect genuine  stars to
have  FWHM $\simeq  2.7-2.8$ pixels  at  our final  drizzled scale  of
$0.035''$/px.   Equally important for  our purposes,  as will  be seen
below, there is  almost no overlap between the  stars and the measured
sizes of  the GCs (that is,  their intrinsic sizes  convolved with the
PSF), with a clear gap between the true stars and the smallest GCs.

To  pick out the  stars, we  used the  SExtractor software  (Bertin \&
Arnouts 1996) to make preliminary measurements of the half-light radii
and FWHMs of  all 2000+ objects on our  detection list, including GCs,
stars, and  a few small  background galaxies.  These scale  sizes were
plotted against the SE  total aperture magnitude.  On these diagnostic
graphs, the stars then show up as an easily identified, narrow (though
thinly populated)  sequence clearly separated  from the GCs  and other
objects.  See,  for example,  Figure 2 of  Harris (2009) for  the same
approach.  We selected the $30$  brightest of these and inspected them
individually  on  the images,  rejecting  any  with faint  companions,
nearby bad pixels, or other anomalies;  one bright star that was at or
near saturation  was also rejected.   Our final starlist has  17 stars
distributed uniformly across the ACS field, from which the PSF is then
constructed.   To   minimize  any  later   concerns  about  systematic
differences  between the  F606W and  F814W images  in the  measured GC
parameters,  we used  exactly  the same  17  stars to  derive the  PSF
separately in each filter.  The  final PSF profiles have FWHMs of 2.75
pixels (F606W) and 2.78 pixels (F814W). 

We create the PSF for each  filter by using the standard {\sc daophot}
routines within  {\sc pyraf}. We  perform photometry of  the candidate
stars with {\sc phot} and then run {\sc pstselect} to select the stars
to be used to build the PSF  by the task {\sc psf}. The output of {\sc
  psf} is  a luminosity weighted  PSF with a  radius of 79  pixels. We
ensured  that  all  stars  used  to  calculate  the  PSF  had  similar
luminosity and thus we avoid  any strong bias in favor of particularly
bright stars.  We  use {\sc seepsf} to subsample our  PSF to the pixel
size required by {\sc ishape},  i.e.  ten times smaller than the image
pixel size (Larsen 1999).

%---------------------------------------------------------------------

\section{Structural Parameters}

\subsection{Effective Radius}

We  use the  software  {\sc ishape}  developed  by S\o  ren Larsen  to
calculate  the  effective  radius,  ellipticity,  and  position  angle
$\theta$  for each  individual cluster  on our  images  (Larsen 1999).
{\sc  ishape} convolves an  analytical model  of the  cluster profile,
with  the  PSF and  finds  the  best fit  to  the  data by  performing
iterative  adjustments to  the cluster  profile FWHM.   The analytical
model used to fit the clusters is chosen by the user from a predefined
list e.  g.  Gaussian, King,  Sersic, see Larsen (1999).  We use model
King  profiles (King, 1962)  with a  concentration parameter  of c=30,
where $c=\frac{r_t}{r_c}$,  the ratio of  the tidal radius  divided by
the core  radius.  This c-value  accurately represents the  average of
real  globular  clusters (Harris  1996,  Larsen  1999,  Larsen et  al.
2001).  By contrast, Waters et al.  (2009) used c=10, a value near the
{\it minimum} for real globular clusters rather than the mean.

The measurement of the effective radius carried out by {\sc ishape} is
robust  and nearly  independent of  the  fitting function.   We set  a
fitting radius  of six pixels and  we convert our  measurements of the
cluster FWHM  into the effective radius by  following the prescription
of Larsen (1999). Namely, we  convert the {\sc ishape} internal radius
for  the FWHM,  which is  model  dependent, into  a model  independent
quantity $r_h =  1.48\times FWHM$, where 1.48 is  the respective value
for c=30.  We  obtain thus the cluster size in  pixels, we multiply by
the  scale, and  then obtain  the  cluster $r_h$  in arcseconds.   All
clusters considered in the following analysis have an effective radius
measured with  a signal-to-noise ratio $>50$. The  {\sc ishape} Manual
(Larsen 2008) has a tentative formula to take into account the effects
of ellipticity  on determining  the effective radius.   Strictly, {\sc
  ishape} measures the  $r_h$ along the major axis  of the profile and
in theory a small correction should  be made to derive $r_h$ for a non
circular  cluster (Larsen  2008).   Given that  most  clusters of  our
sample are nearly  circular we quote the direct  values of $r_h$ given
by {\sc  ishape} without  further corrections for  the effects  of the
ellipticity during the fitting process.  Details on the sensitivity of
our measurements to  cluster ellipticity and tests of  the validity of
our  assumptions are  given in  Harris (2009).   Results are  shown in
Figures 2 and 3.

An  extensive  set  of  tests  and  simulations  to  characterize  the
uncertainties associated  with {\sc ishape}  measurements is presented
in Harris (2009).  In this work  it is clearly shown that {\sc ishape}
derives accurate values for  cluster effective radius, ellipticity and
position angle  when S/N$>$50, which is  the cut-off that  we adopt in
this study.  The main sources of error in the size measurements can be
summarized  in three  components: the  {\sc ishape}  fitting procedure
itself,  the  uncertainty in  the  true  values  of the  concentration
parameter c, and  the error associated with the size  of the PSF.  The
random uncertainty per object taking into account the three sources of
uncertainty cited  above is $\sigma _{r_h}=  \pm 0.006\arcsec$ (Harris
2009) for his BCG data  sample. For comparison, the rms scatter around
the 1:1  line in Figure  2 is $\pm  0.004\arcsec$; this should  give a
reasonable estimate  of our internal measurement  uncertainty since it
is the direct  comparison of two independent measurements  of the same
objects in different filters. The equivalent uncertainty for $r_h$ is
then $\pm 0.3$ pc.

The values for the linear effective  radii that we derived are in good
agreement with the  values of Milky Way globular  clusters.  The Milky
Way has  142 clusters  with known effective  radius, out of  these 142
clusters  114 (80\%)  have  an effective  radius  between 1  and 6  pc
(Harris 1996). It can be seen in Figure 2 and the histograms in Figure
3 that most M87 clusters also have an effective radius between 1 and 6
parsecs and  peak at very much the  same radius ($\simeq 2.5$  pc ) as
the Milky Way.

It  is  intriguing  that  clusters  belonging  to  two  galaxies  with
different Hubble type have  the same size distribution.  Forbes (2002)
plotted side by side the size distribution of GCS for five galaxies of
completely different  Hubble type: cD,  gE, S0, Sa,  Sbc.  Strikingly,
the size distributions of the overall GCS, and of both sup-populations
are very  similar, indicating that the  physical processes determining
the size of globular clusters  and the size difference between the two
sub-populations  are  independent of  galaxy  type.   Whitmore et  al.
(2007) suggest a universal initial mass function for clusters followed
by an  evolution dominated by  internal dynamics. The host  galaxy has
only a secondary effect on cluster size.

We   recover  the   well  documented   size  difference   between  red
(metal-rich) and  blue (metal-poor)  clusters, i.e. blue  clusters are
24\% larger  than red  ones in the  M87 GCS  (a result first  noted by
Kundu et al. 1998), as discussed further below. 

We also find a hint of  a size difference versus wavelength such that,
the  cluster size  measured  in  $F606W$ is  very  slightly larger  on
average than in the $F814W$.  For all clusters we calculated the ratio
$\frac{r_{h}F606W}{r_{h}F814W}$; the  histogram of the  values of this
ratio is plotted  in Figure 4.  Of the 1896  clusters which are within
our  high confidence  range 61\%  of them  have $r_{h}  F606W  \ge r_h
F814W$.  The  median ratio, near the  peak of the  histogram, is $1.02
\pm 0.006$ and the standard  deviation is $0.24$.  We rejected objects
at more than  $3\sigma$ level as outliers. We are  well aware that the
mean  difference  may have  arisen  from  small residual  systematics.
However we have attempted to  minimize any such systematics as much as
possible,  and we  should  note  that the  very  small raw  difference
between the two PSF sizes noted above is not large enough by itself to
produce  the  difference we  see,  particularly  for bigger  clusters.
However, if  the difference in mean  $r_h$ between the  two filters is
physically real, we suggest that  we may be seeing the visible results
of \emph{mass segregation}.  Dynamical  evolution of the population of
stars  within each  cluster drives  the less  massive stars  to larger
radii  and  the  more  massive  ones inward  toward  the  core,  while
maintaining  a   nearly  constant  $r_h$.   Over   time,  a  secondary
observational effect  is that the  cluster effective radius  will look
progressively smaller  in redder bandpasses  that systematically favor
the light from the massive red-giant and subgiant stars, compared with
the  bluer,  lighter  upper-main-sequence  stars  that  preferentially
populate the outskirts.  To gauge  the expected size of the effect, we
have used the  predicted $r_h$ values in the $B,  V, I$ bandpasses for
model  globular clusters  with  an initial  mass  of $10^5  M_{\odot}$
evolved  through  an  advanced   N-body  code  (Hurley  2009,  private
communication;  see also  Hurley  et  al. 2008).   After  10-12 Gy  of
dynamical  evolution, the  model clusters  have measured  $r_h$ values
that are 5\%  larger in $V$ than in $I$.  Both  the direction and size
of the effect are very close to the mean offset that we see in the M87
system.  We view  our result as only suggestive, but  it may point the
way  to a  valuable  new test  of  our understanding  of GC  dynamical
evolution. We will further investigate this effect with more data that
we are currently analyzing.

%---------------------------------------------------------------------

\subsection{Ellipticity}

{\sc  ishape} measures  the ratio  of  minor/major axis  and thus  the
ellipticity of each cluster.  In  Figure 5 we plot the values obtained
with {\sc ishape} for the  ratio of minor/major axis for 1767 clusters
of our dataset.  Similarly to  the Milky Way globular clusters (Harris
1996) most  M87 clusters  are roughly spherical,  i.e.  for  more than
55\% of clusters in our sample the ratio minor/major $>$ 0.8. A normal
two sample  Kolmogorov-Smirnov test shows  that the M87 and  Milky Way
distributions are formally different at high statistical significance.
However, these two samples were measured in completely different ways:
the Milky  Way clusters  are orders of  magnitude better  resolved and
thus  their ellipticities  are much  better determined  and physically
meaningful.   Figure 5  shows  that, roughly,  M87  clusters have  the
expected range of small ellipticity.

%---------------------------------------------------------------------

\subsection{Effective radius as a function of galactocentric distance}
 
Hodge  (1960,  1962) measured  the  sizes  of  Large Magellanic  Cloud
globular  clusters  using  data   taken  with  the  ADH  Baker-Schmidt
telescope in South Africa. Despite the fact that Hodge (1962) observed 
a projected distribution of the cluster population instead of the real
spatial  distribution,  he first  recognized  the correlation  between
cluster  size and  galactocentric radius  ($R_{gc}$), i.   e.  cluster
size increases with  increasing distance to the center  of the galaxy.
Hodge (1962) postulated the tidal  effects of the parent galaxy on the
cluster as  the most probable  explanation for this effect.   Based on
data for Milky Way globular clusters, and therefore free of projection
bias, van  den Bergh  et al.  (1991)  find a relation  between cluster
effective radius  and true  3D galactocentric distance  of $r_h\propto
\sqrt{R_{gc}}$. The same dependence of cluster size and galactocentric
distance has  been well  documented in NGC5128  (Hesser et  al.  1984,
G\'omez \& Woodley  2007), NGC4594 (Spitler et al.   2006), and in six
other giant ellipticals (Harris 2009).  For these galaxies the general
trend is  for $r_h$ to increase roughly  as $r_h\sim R_{gc}^{0.1-0.2}$
where $R_{gc}$ is the projected distance.

In  Figure 6  we  plot  the effective  radius  $r_h$ versus  projected
galactocentric distance $R_{gc}$.  Our observations target the central
regions of M87  where tidal forces exerted by  the galaxy are expected
to have the strongest effect  limiting the size of clusters (Hesser et
al  1984). For  $R_{gc} \le$  5 kpc  the mean  cluster size  is nearly
constant, but then begins to  increase gradually, out to the limits of
our data  at $R_{gc}=11$  kpc. Kundu et  al. (1999) failed  to observe
this trend given their smaller WFPC2 field size.

In  order to  demonstrate  that  the increase  in  effective radii  of
clusters with $R_{gc}$ is real, we  performed a test with the stars on
the field.  We selected all  the objects that {\sc ishape} returned as
having  FWHM=0, that  is the  ones  that are  by definition  starlike.
There were  42 such stars, we  then measured with  {\sc imexamine} the
FWHM of these point sources.  These  stars can also be clearly seen in
Figure  2, 3  \&  6  as the  sources  around the  zero  value for  the
effective radius, and  are also apparent in Figure 3  as the first bin
of the histogram  i.e. $r_h=0$.  We find that the  FWHM of these point
sources is  consistent with the value  of the PSF FWHM  of 2.7 pixels,
and more  importantly does not  show a dependence  with galactocentric
radius (see  Figure 7 top  panel).  We also  plot in Figure  7 (bottom
panel)  the residual  of the  PSF with  unresolved sources  across the
detector that is, the ``size'' of  each star as returned by the ISHAPE
fit. This  additional test to the  accuracy of the PSF  shows that the
residuals  are minor and  do not  show any  specific trend  across the
detector.

%---------------------------------------------------------------------

\section{Photometry}

The photometry  section of this paper  is placed at the  end to better
reflect the  steps we followed  in our analysis. Indeed,  we calibrate
our  photometry {\it  after} we  obtain the  effective radius  for our
clusters as follows.

We determine  the magnitudes on the  output images of  APSIS using the
expression:

\begin{displaymath}
m_{VEGA}  =  - 2.5log(FLUX \times CORRECTION)+2.5log(EXPTIME) +ZP
\end{displaymath} 

where {\sc flux} is the number of counts within a circular aperture of
5 pixels (0.175$\arcsec$) in radius.  {\sc correction} is the aperture
correction  factor that  we  apply  and that  we  describe below.  {\sc
  exptime} is  the total exposure  time for each  image and ZP  is the
zeropoint in the VEGAMAG system.

Our data  was taken before the  failure of the ACS  Side 1 electronics
and following temperature change of  the camera in 2006 June (Sirianni
et al. 2006).  Our photometric  zeropoints are obtained from the STScI
website: $ZP_{F606W}=26.420$ and $ZP_{F814W}=25.536$.

To  calculate the  aperture correction  we used  the  procedure below.
Using  the {\sc  ishape} tasks  {\sc  mkcmppsf} and  {\sc mksynth}  we
create  model images of  synthetic globular  clusters.  We  make these
synthetic clusters with different  profiles of varying intrinsic FWHM,
and convolve  them with the  PSF.  In order  to recreate the  sizes we
obtained for the effective radius,  we built clusters with FWHM from 0
to 3  pixels by  steps of 0.02  pixels. These synthetic  clusters have
different curves of growth that we  plot in Figure 8.  With the curves
of growth  for these  synthetic point sources,  of known FWHM,  we can
perform  an accurate aperture  correction by  measuring the  offset in
flux (or  magnitude) for each cluster  size to a stellar  PSF. Once we
determine the  offset to  a stellar source  we can apply  the aperture
correction of Sirianni  et al.  2005.  Given that  the Sirianni et al.
(2005) tables do not  give an aperture correction for r=0.175$\arcsec$
we use  an interpolated  value, i.e. 0.8205  for F606W and  0.7985 for
F814W.   With the  above method  we  obviate the  objections of  Kundu
(2008) related to  the use of a fixed aperture to  measure the flux of
clusters with different sizes.

\subsection{Color-Magnitude Diagram}

We  present our  final color-magnitude  diagram for  the  M87 globular
clusters in Figure 9. We plot the F814W magnitudes in the y-axis since
these are  a better  tracer of  the cluster stellar  mass. The  CMD in
Figure   9  is   consistent  with   previously  published   M87  CMDs,
particularly with the CMD derived  by Larsen et al. (2001) using WFPC2
data and a similar pair of filters i.e.  the F555W and the F814W.

The sharpness  of the M87 bimodality  is striking and  is more clearly
defined than in previous work (e.  g. Larsen et al.  2001, Peng et al.
2006) because  of the much  higher S/N of  this deep dataset.   In the
magnitude range $F814W\simeq  22.0 - 22.5$ we note  a "bridge" between
the normal blue and red  sequences, with many more clusters than usual
at  intermediate colors.   These do  not seem  to lie  in  any special
location, nor to have scale  sizes different from other clusters. They
may   be  genuinely   intermediate-metallicity  objects.    Given  the
age/metallicity  degeneracy  for  old  clusters  they  might  also  be
metal-rich  but younger  by  a  few Gigayears  than  the red  sequence
clusters. Further discussion is given in Peng et al. (2009).

The CMD presented  here differs from the CMD derived  by Waters et al.
(2009). One  of the  most evident differences  is the presence  in the
Waters et al.  (2009) CMD  of bright ($I<22$) clusters that are either
very blue ($V-I < 0.9$)  or very red ($V-I>1.3$). The equivalent areas
of  our CMD  are completely  devoid of  clusters, suggesting  that the
random errors  in the  magnitudes calculated from  the curve-of-growth
procedure are  low. A  detailed study of  the CMD  for the M87  GCS is
presented by Peng et al. (2009).

%---------------------------------------------------------------------

\subsection{Effective radius versus magnitude}

In  Figure 10  where we  plot  magnitude versus  effective radius  for
metal-poor and metal-rich clusters,  the boundary between red and blue
clusters is set at F814W-F606W $>$  or $<$ than 0.8. In this figure we
can see  how metal  poor clusters have  larger $r_h$ than  their metal
rich counterparts.  Red  clusters have a median $r_h$  of 2.1 pc while
for blue clusters the median $r_h$  is 2.6 pc.  Blue clusters are thus
on average 24\% larger than  red clusters, this offset is in excellent
agreement with  the findings of  previous studies using  large cluster
samples in other galaxies (e. g.   Larsen et al.  2001, Spitler et al.
2006,  Harris 2009). The  $r_h$ median  value is  not affected  by the
presence of clusters with $r_h\sim 0$ visible at the bottom of the two
graphs of Figure 10.  This size difference between sub-populations was
explained by Jord\'an (2004) as the result of mass segregation and the
dependence   of  main-sequence   lifetimes  on   metallicity.  Another
possibility is  that this difference reflects  conditions of formation
(Harris, 2009).  We do not  observe any clear trends of increasing (or
decreasing) $r_h$ with magnitude.

One clear  fact evident in Figure  10 is the unfilled  space between 0
and $\sim$1.5 pc in both plots at all magnitudes except the faint end.
This gap  between the  unresolved stars and  the smallest  clusters is
also visible in Figures 2, 3,  and 6. It appears as most clusters have
a minimum  effective radius.   This gap in  the $r_h$  distribution is
also present in previous studies of  M87 (Kundu et al. 1999, Larsen et
al.  2001). Of the Cen A and Milky Way clusters none has been reported
with  an effective  radius of  less than  1pc (G\'omez  et  al.  2006,
G\'omez \& Woodley 2007). 

An  explanation for  the existence  of this  is gap  is the  fact that
smaller clusters have a smaller chance of survival. McLaughlin \& Fall
(2008) study  the relation of  the globular cluster mass  function and
the cluster half-mass density ($\rho_h$). In their study these authors
normalize the disruption time of  a globular cluster to the relaxation
time and find that the  evaporation rate $\mu_{ev}$ is proportional to
$\rho_h^{1/2}$.   Using  their   relation  between  $\rho_h$  and  the
effective  radius we  can write  $\mu_{ev} \propto  r_h^{-3/2}$.  This
means  that smaller clusters  evaporate much  faster than  bigger ones
(Fall  \&  Rees  1977,  1985).   The evaporation  rate  determined  by
McLaughlin \& Fall (2008)  only takes into account internal relaxation
effects.  The  influence of an  external potential on  the dissolution
timescale is  examined by Gieles  \& Baumgardt (2008).   These authors
conclude that  in the presence of  a tidal field  the dissolution time
scales with  the number of  stars in the  cluster $N$ and  the angular
frequency   $\omega$   of   the    cluster   in   the   host   galaxy:
$t_{dis}=\frac{N^{0.65}}{\omega}$.

%---------------------------------------------------------------------

\section{Comparison with a recent study}

A recent paper by Waters et  al. (2009) analyzed the same dataset used
here but aimed at discussing  the mass/metallicity relation in the GCS
and its color bimodality.  Waters  et al.  (2009) do not make specific
conclusions on structural parameters  for the clusters themselves.  We
note below several  steps in our analysis of  this dataset that differ
from their work.

A first important difference in  our data reduction concerns the final
combined image  that we use  to measure the structural  parameters. In
this  study  the  final  image  is  subsampled to  a  pixel  scale  of
0.035\arcsec  taking   thus  full   advantage  of  the   dithered  raw
observations.  Waters et al.  (2009)  keep the ACS camera native pixel
size of 0.05\arcsec.  The PSF is also computed in a different fashion:
Waters  et al.  (2009)  use the  software of  Anderson \&  King (2006)
while in this  work we create a purely empirical  PSF using a sequence
of tasks described in section 4.

The magnitudes and radii of  $\sim$ 2000 clusters, each with different
size  cannot be  measured with  either standard  PSF-fitting  or fixed
aperture  techniques. In order  to derive  proper magnitudes  both our
work and Waters  et al.  (2009) compute flux  corrections based on the
size of  each object.   In this study  we calculate the  magnitudes by
scaling the flux  that we measure at a 5 pixel  radius to large radius
using the curve of growth of each object appropriate to its particular
effective  radius.   The details  of  this  method  are given  in  the
photometry  section. Waters et  al. (2009)  instead use  the magnitude
difference between  a radius of 4 pixels  and a radius of  2 pixels as
the main  parameter to  estimate the cluster  size and  their aperture
correction.  Lastly, our  measurements of cluster size is  done with a
different  code,  {\sc  ishape},   and  using  a  fiducial  King-model
concentration  $r_t/r_c=30$ that  is more  representative of  the true
mean for  globular clusters than  their adopted ratio  of $r_t/r_c=10$
(see section 5 above).

When  compared with  Waters et  al.  (2009)  some of  our  methods are
obvious  improvements,   such  as  obtaining  an   image  with  higher
resolution,  while other  differences are  an independent  approach to
measure the  same quantity,  such as the  use of an  alternate fitting
profile code.

%---------------------------------------------------------------------

\section{Conclusions}

With extremely deep HST/ACS data in  V and I, we have obtained precise
measurements  of  the  scale  radii  $r_h$ for  $\sim$  2000  globular
clusters  in M87.   To first  order, their  size  distribution closely
resembles  the Milky  Way  GCS. We  find  that the  mean cluster  size
depends  noticeably on both  metallicity and  galactocentric distance.
Metal rich clusters are 24\% larger than the metal-poor subpopulation.
The mean $r_h$, for all clusters, begins to increase with $R_{gc}$ for
$R_{gc}\ge$ 6 kpc.

A new, but very tentative, finding  of this work is that the effective
radius  of  these  old  globular  clusters  may  depend  on  bandpass,
appearing slightly  larger at bluer wavelengths.  This  effect is just
on  the  margin of  being  statistically  significant,  but if  it  is
physically  real, it  may point  to  the effects  of mass  segregation
within the clusters.  Future  studies from other, perhaps more nearby,
galaxies should be able to test this idea more strongly.

The  measurement   of  structural  parameters   of  globular  clusters
belonging  to  galaxies  of  different  Hubble  type  and  located  in
different  environments  should  reveal  in the  future  any  existent
correlation  between galaxy  type and  cluster sizes  (G\'omez  et al.
2006).

%---------------------------------------------------------------------
%---------------------------------------------------------------------
%---------------------------------------------------------------------

\acknowledgments

We  are grateful  to Jennifer  Mack,  and Marco  Sirianni (STScI)  for
kindly answering several queries related  to the ACS.  We thank Warren
Hack (STScI)  for helpful discussions  on Multidrizzle. We  also thank
Jarrod Hurley  (Swinburne University,  Australia) for sharing  with us
evolved globular  cluster models.  This  research has made use  of the
NASA Astrophysics Data System Bibliographic services. STSDAS and PyRAF
are  products  of the  Space  Telescope  Science  Institute, which  is
operated by AURA for NASA.
 
%---------------------------------------------------------------------

{\it Facilities:} \facility{HST (ACS)}

%---------------------------------------------------------------------

%---------------------------------------------------------------------
%---------------------------------------------------------------------
%---------------------------------------------------------------------
%---------------------------------------------------------------------

\begin{figure}                            
\plotone{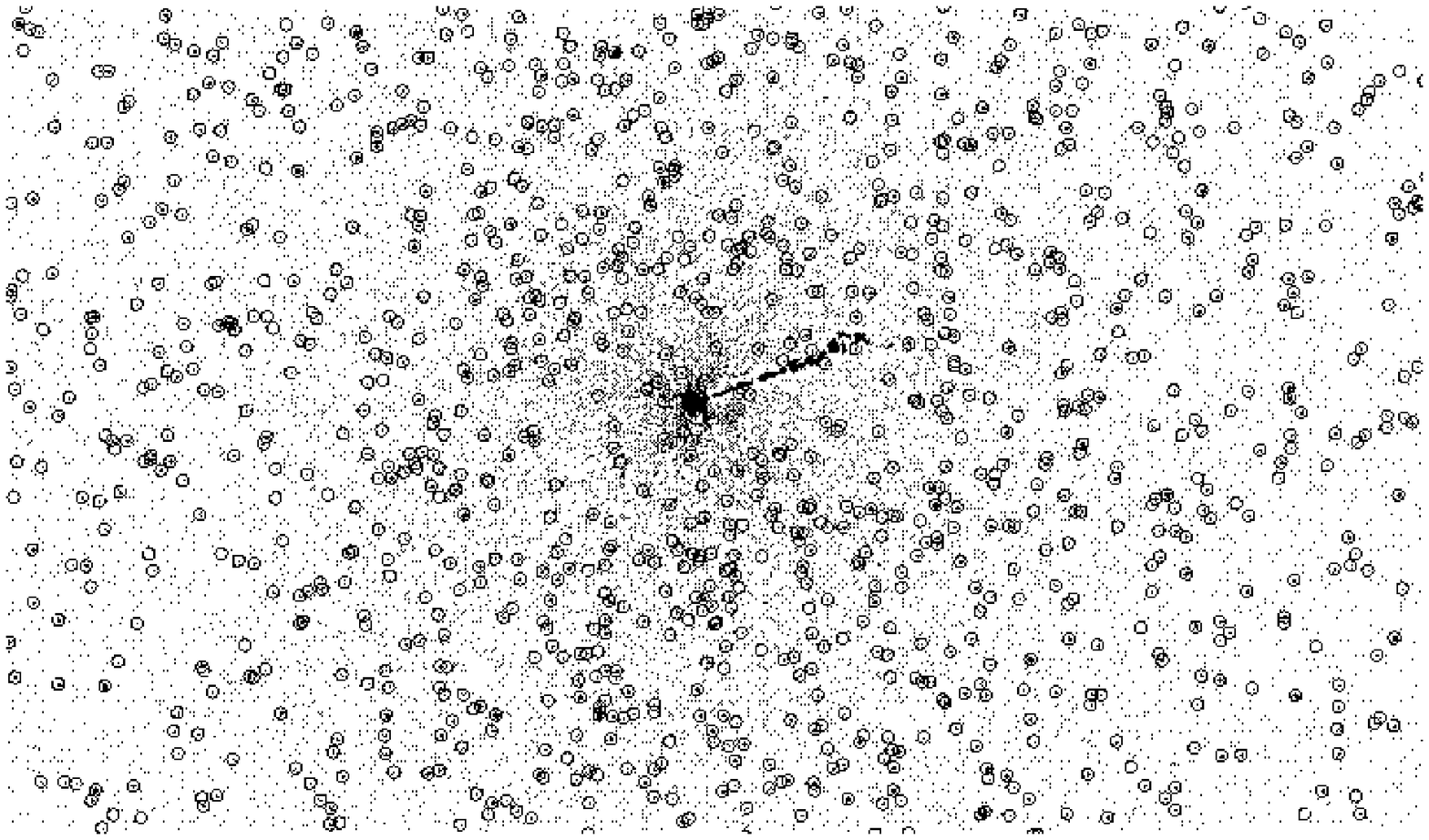}
\caption{Median subtracted image of a  portion of the ACS/WFC field of
  view  of the inner  region of  M87. Detected  clusters are  shown by
  circles, the  prominent plasma jet emanating  from the super-massive
  black hole at the core of  the galaxy is clearly visible. This image
  is $\sim 13$ kpc across, north is up and east is left.\label{fig1}}
\end{figure}

%---------------------------------------------------------------------

\begin{figure}
\plotone{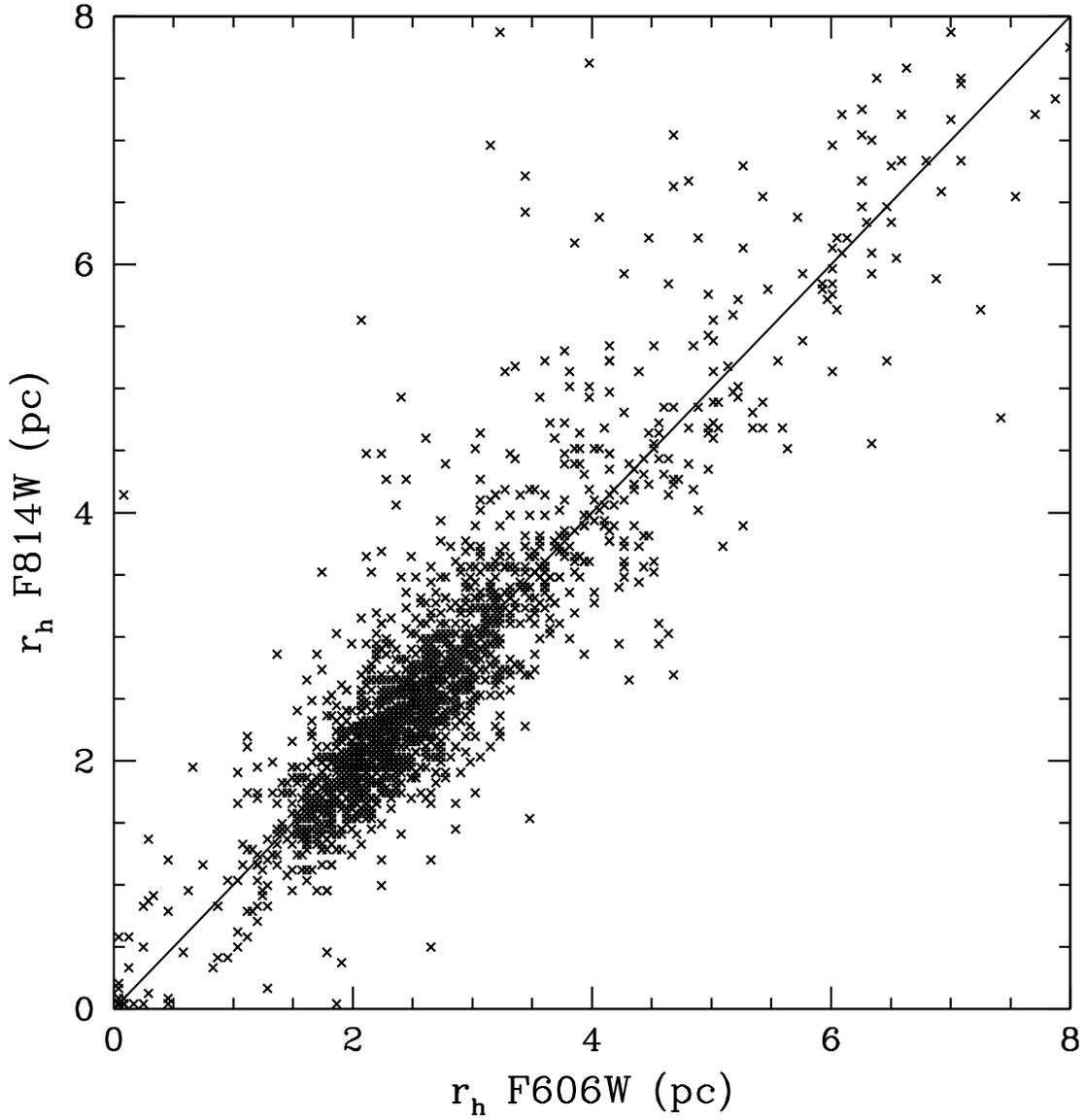}
\caption{Cluster  effective  radius in  F606W  vs.  cluster  effective
  radius in F814W.  Note the  gap between a small number of unresolved
  objects,  clustering around 0,  and the  large majority  of resolved
  clusters grouped around $\sim 2.5$  pc.  The units of this graph are
  parsecs. \label{fig2}}
\end{figure}

%---------------------------------------------------------------------

\begin{figure}
\plotone{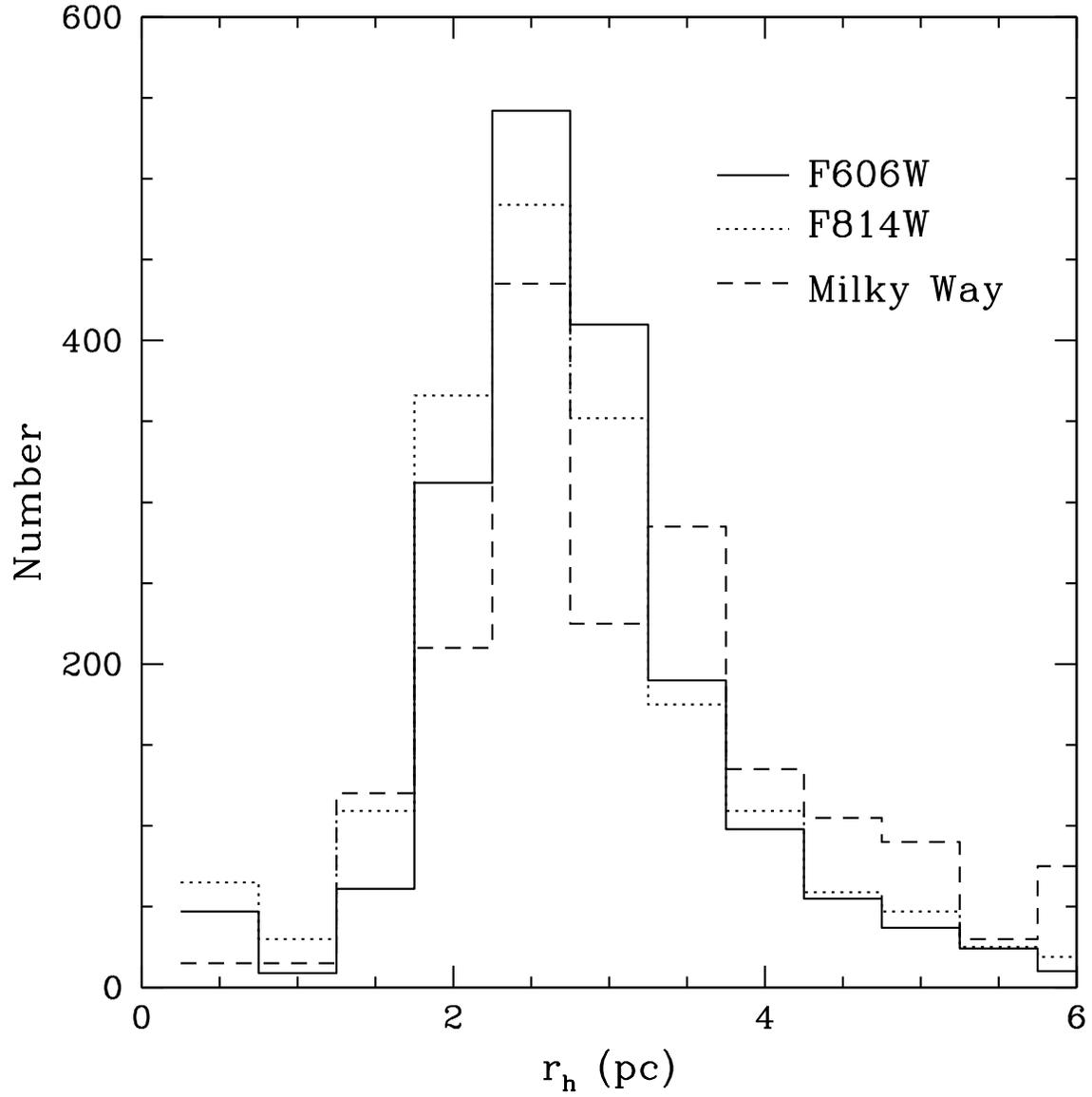}
\caption{Histogram of effective radius in  F606W and F814W.  It can be
  seen  on this  figure  how  the effective radius of most clusters
  lies between 1 and 6 pc. The histogram of the effective radius for
  Milky Way clusters ($\times 15$) is overplotted for comparison.
  \label{fig3}}
\end{figure}
%---------------------------------------------------------------------

\begin{figure}
\plotone{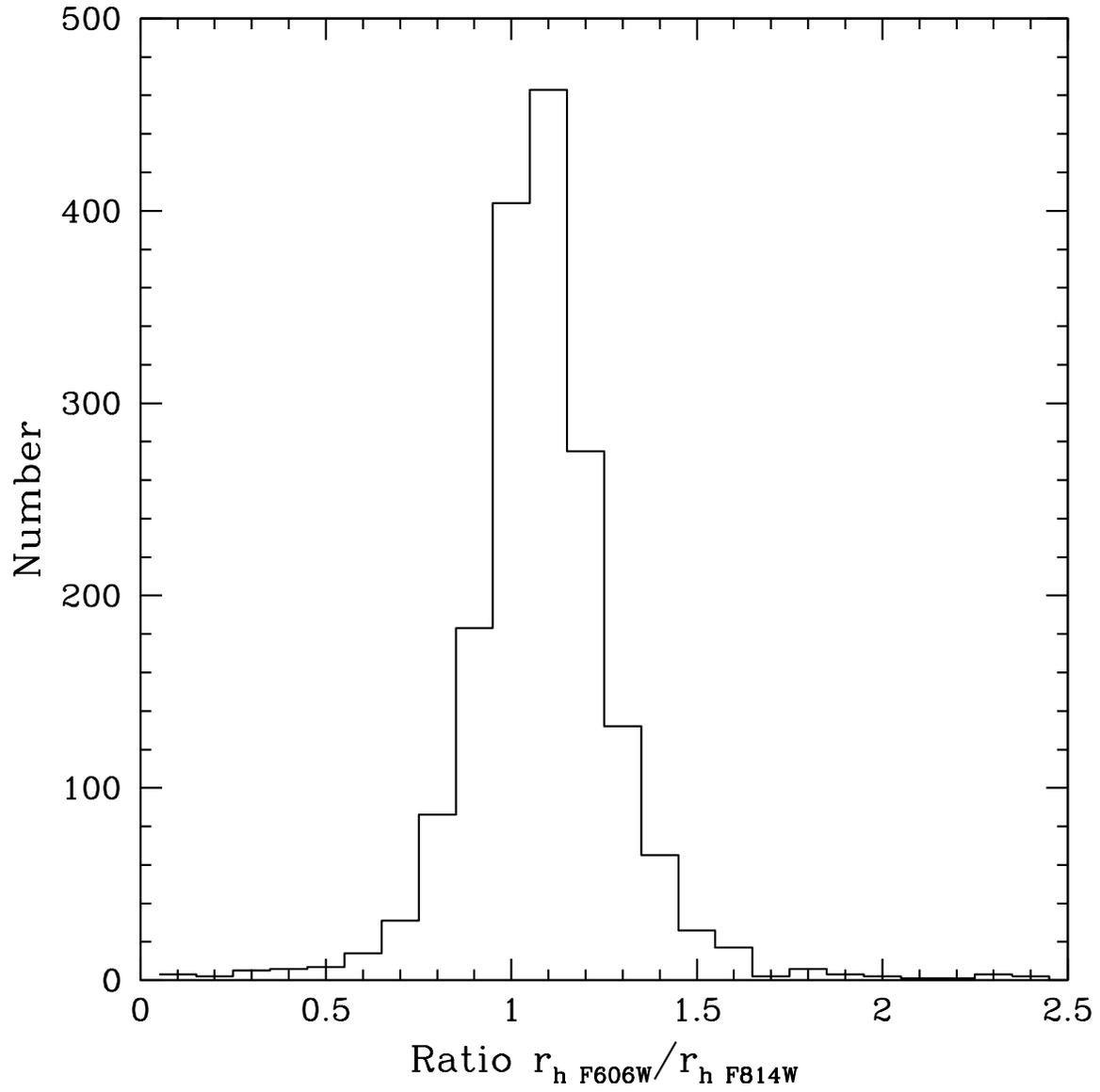}
\caption{Ratio of the  effective radius of M87 clusters  in F606W over
  F814W,  the median  of this  distribution  is 1.02  with a  standard
  deviation of 0.24. \label{fig4}}
\end{figure}

%---------------------------------------------------------------------

\begin{figure}
\plotone{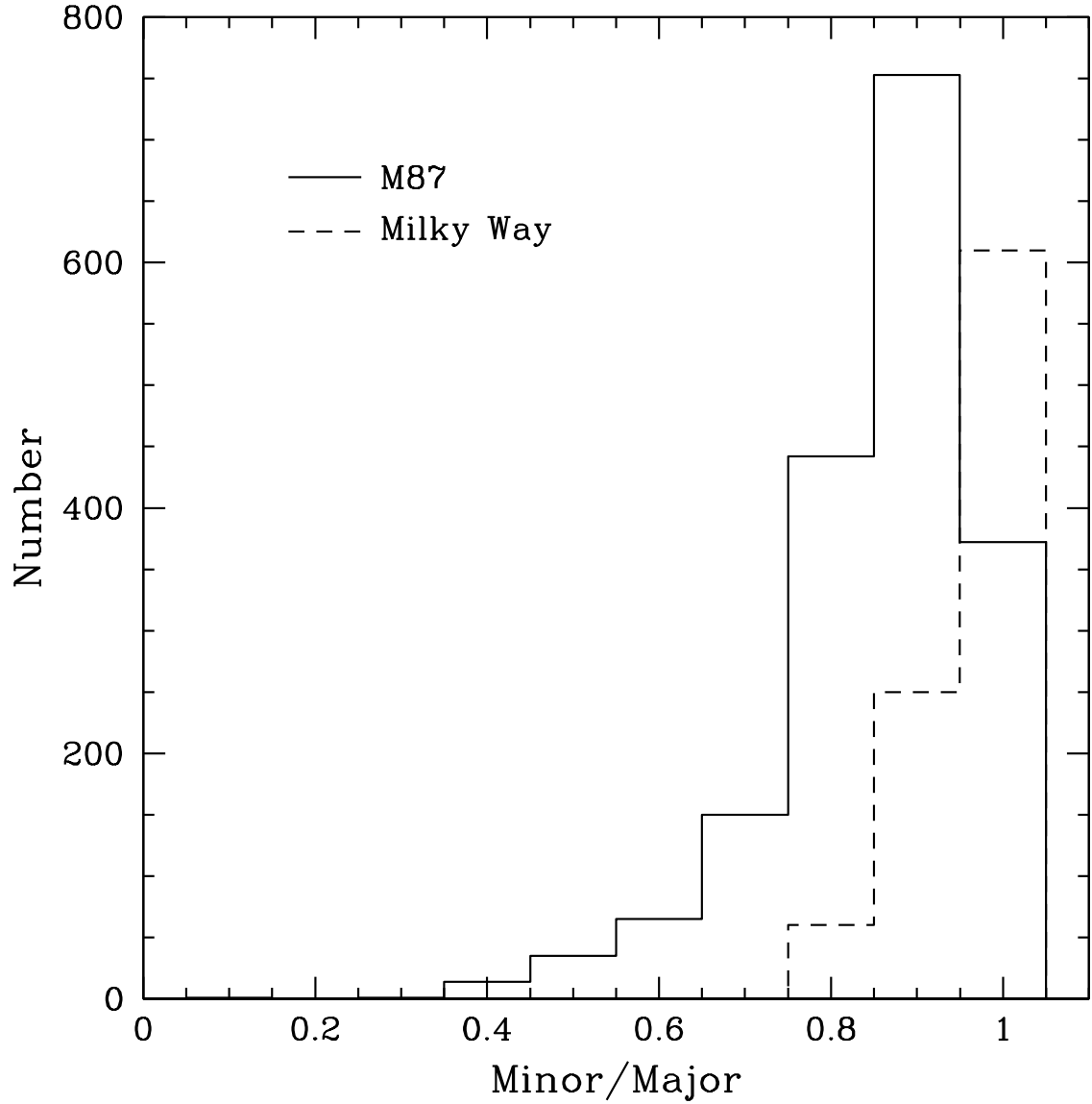}
\caption{Values  for  the  ratio  of  minor/major  axis  for  the  M87
  clusters. Most clusters  are nearly round (e$<$0.2),  though the mean
  ellipticity is slightly larger than in the Milky Way.}
\end{figure}

%---------------------------------------------------------------------

\begin{figure}
\plotone{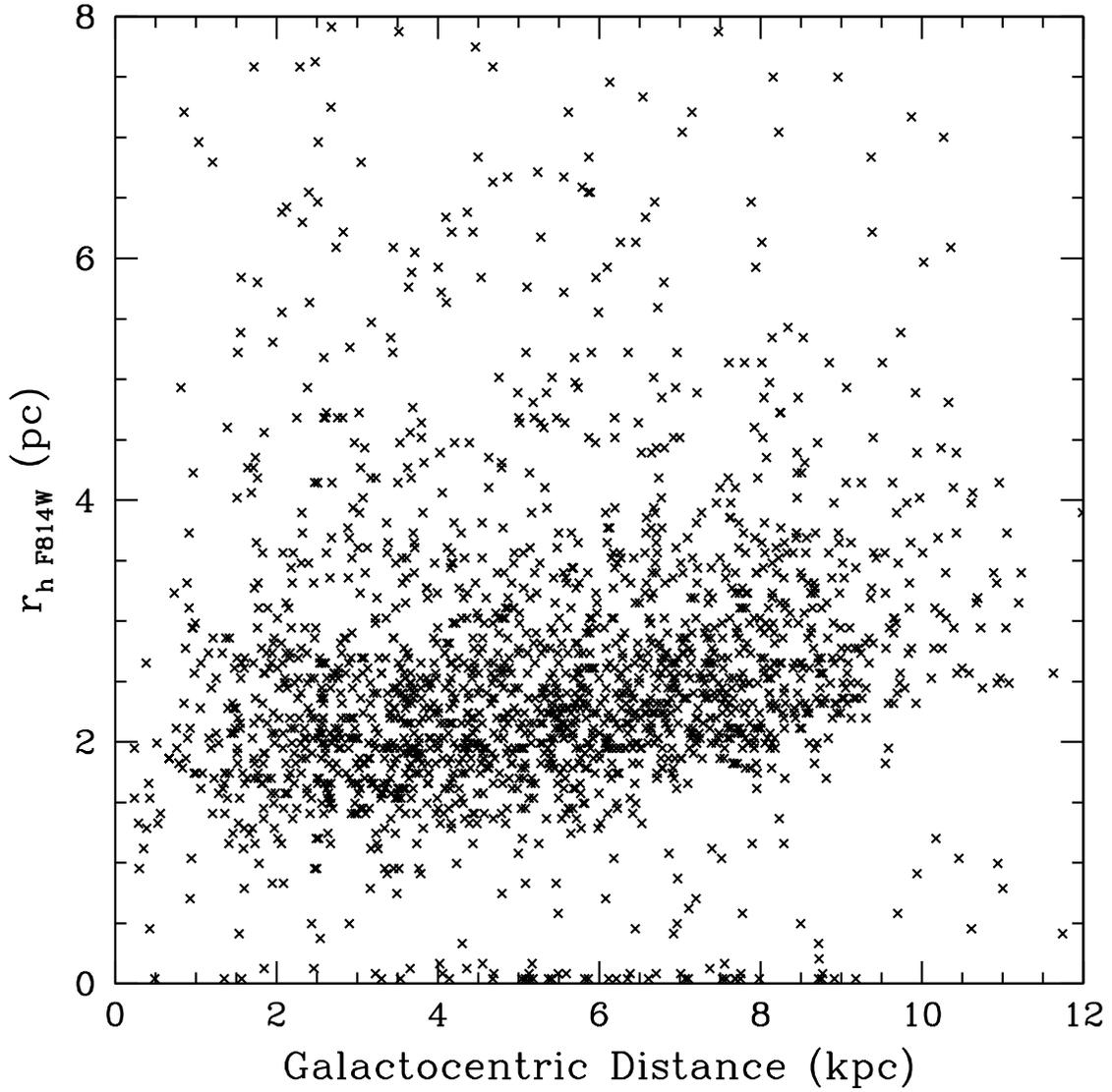}
\caption{Effective radius, in  the F814W filter, versus galactocentric
  distance. Note the small  number of starlike objects scattered along
  the bottom of the graph, clearly separated from the globular cluster
  population.}
\end{figure}
%---------------------------------------------------------------------

\begin{figure}
\plotone{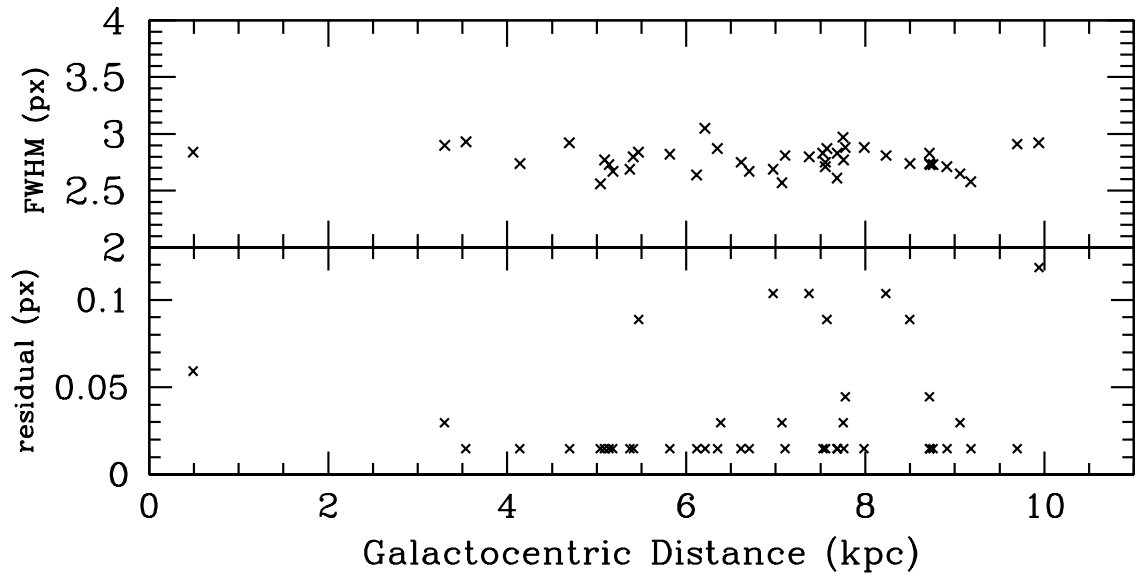}
\caption{FWHM of sources with zero effective radius vs. galactocentric
  distance as  measured with {\sc  imexamine}, top panel.   The bottom
  panel  shows the  residuals  of the  PSF  and the  stars across  the
  detector. We do not see any correlation of PSF size with position on
  the image.}
\end{figure}

%---------------------------------------------------------------------

\begin{figure}
\plotone{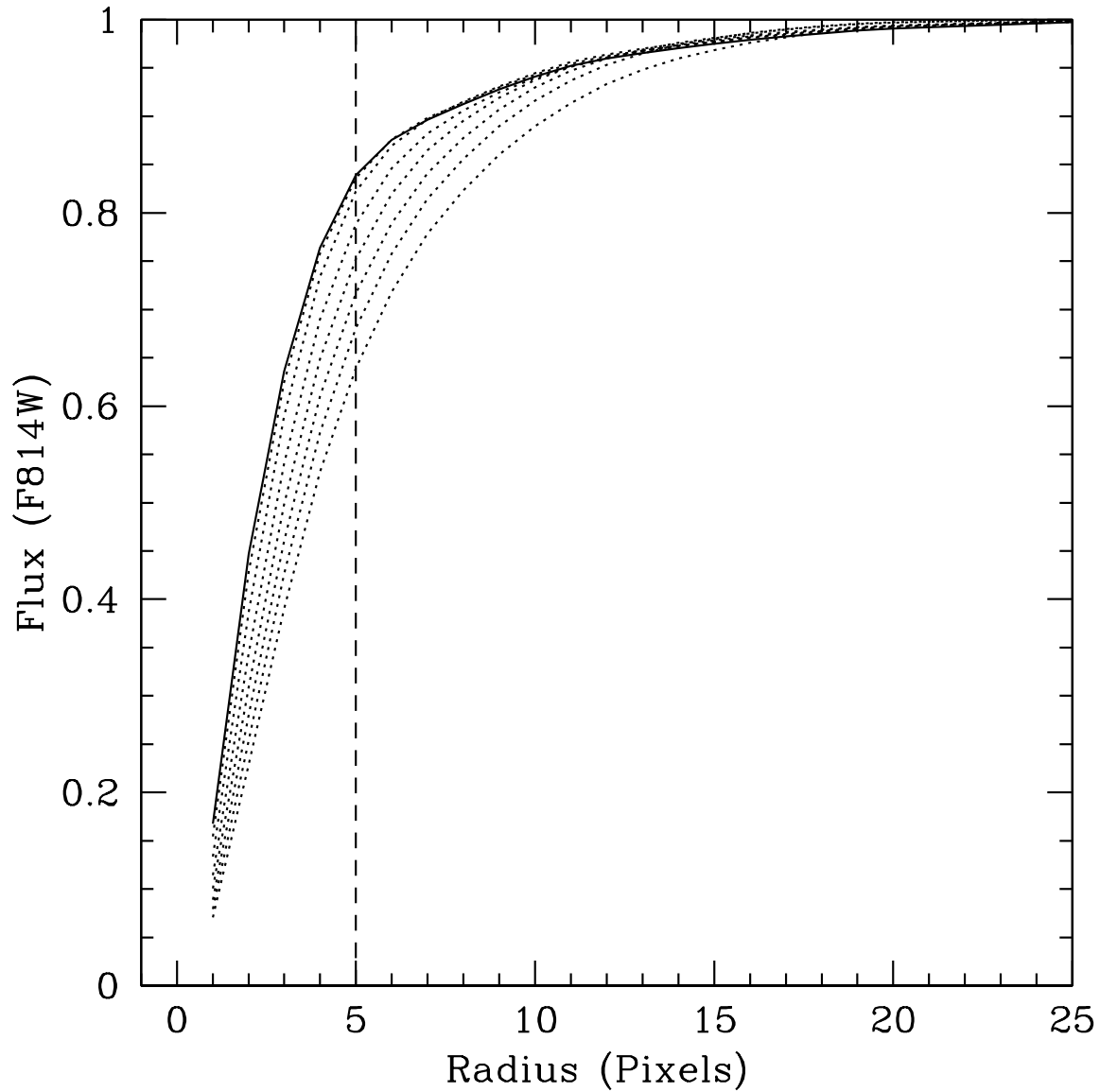}
\caption{Curves of growth for synthetic sources (globular clusters) of
  different FWHM. The solid line  corresponds to a source with FWHM=0,
  i.e. a starlike  object. The dotted lines show  the curves of growth
  of seven sources with FHWM varying from 0 to 1.4 pixels (in steps of
  0.2 pixels)  convolved with the PSF.  The vertical dashed  line at 5
  pixels marks  the radius  of the circular  aperture we used  for our
  photometry.}
\end{figure}

%---------------------------------------------------------------------

\begin{figure}
\plotone{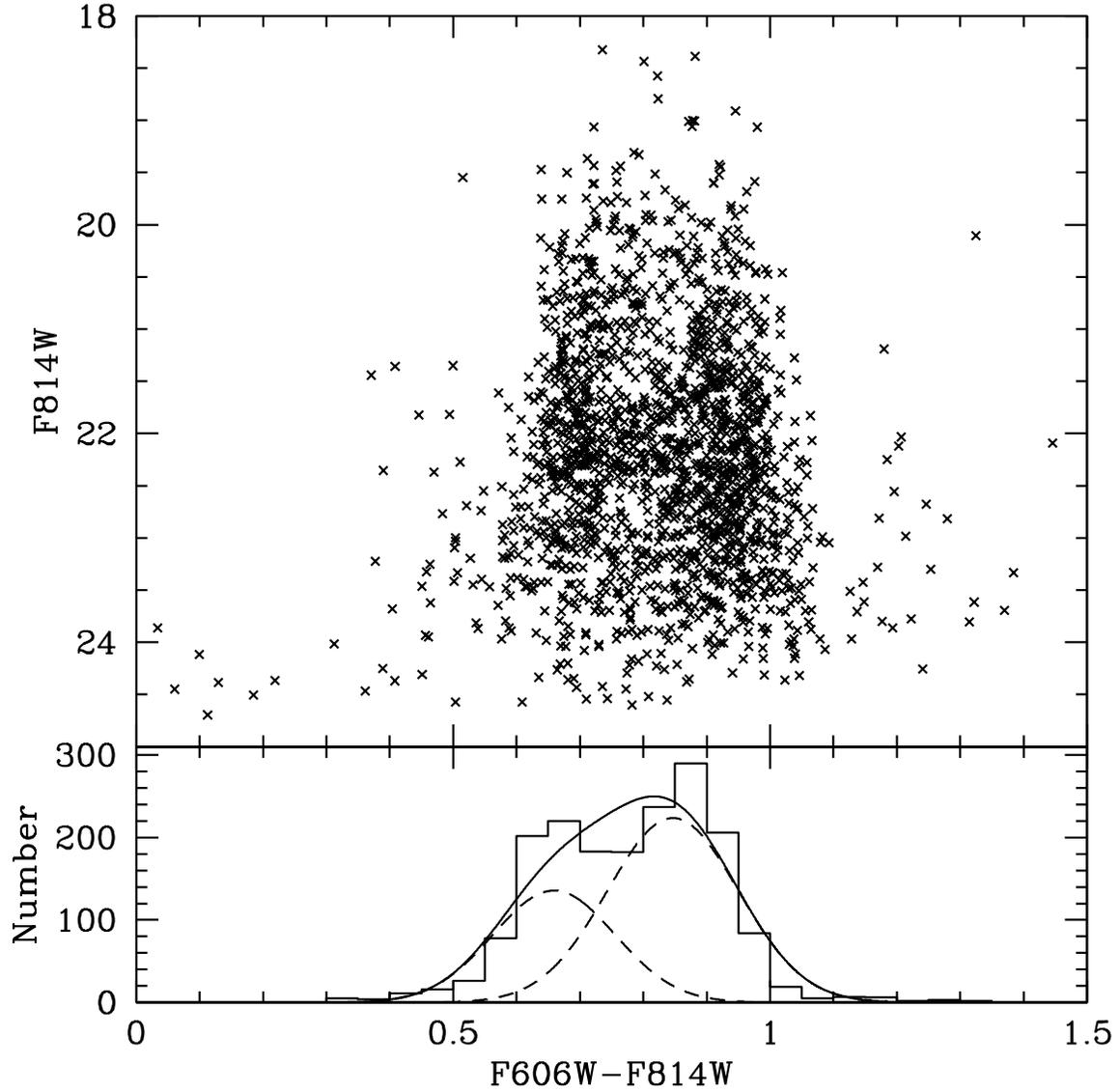}
\caption{Color-magnitude diagram (top) and histogram of cluster colors
  (bottom) for the  M87 Globular Cluster System, the  sharpness of the
  bimodality is  striking. The dashed  lines correspond to  a Gaussian
  fit  to each  subpopulation.  The solid  line  is sum  of these  two
  Gaussians.}
\end{figure}

%---------------------------------------------------------------------

\begin{figure}
\plotone{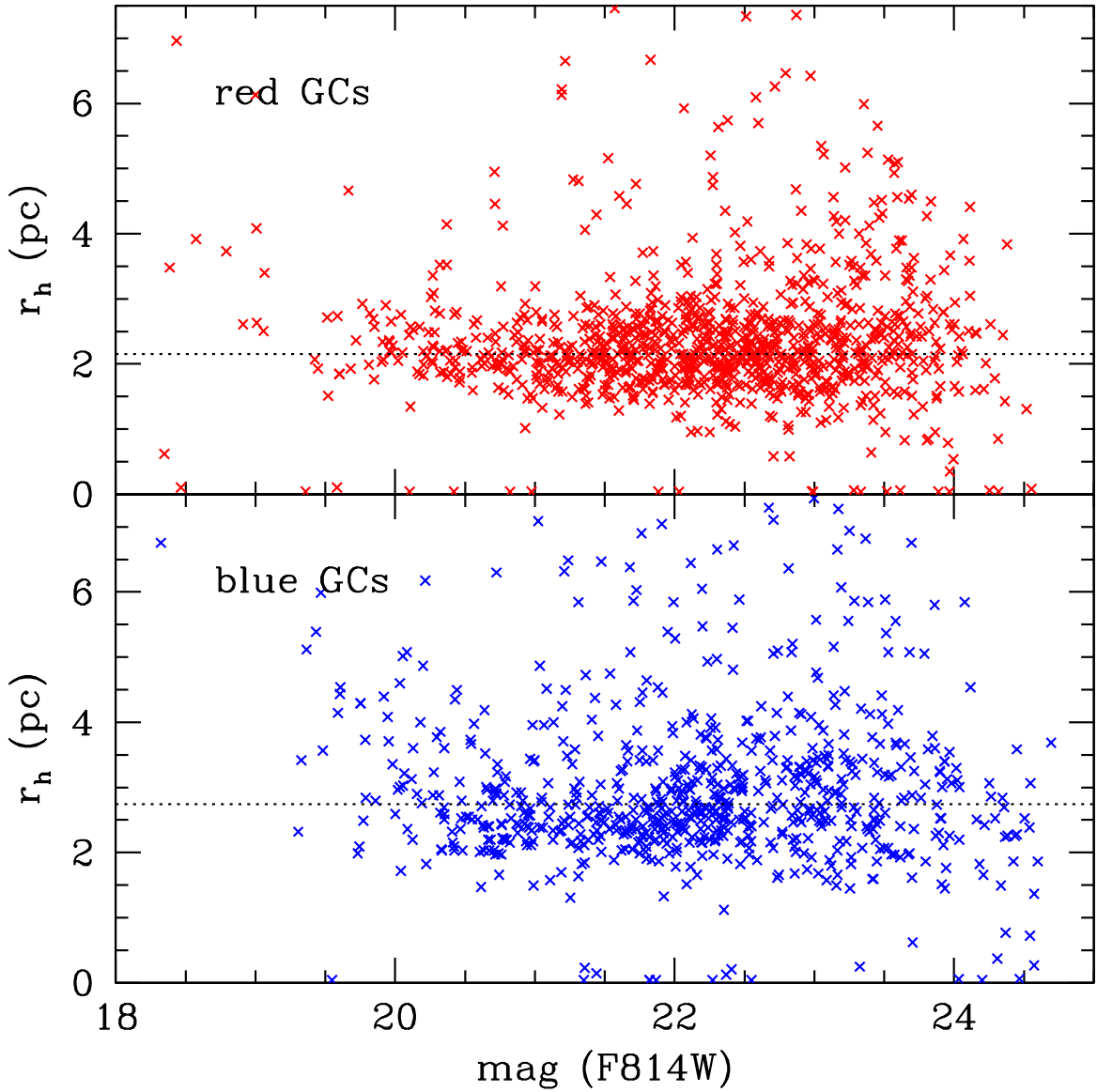}
\caption{Effective radius versus magnitude for the two, red and blue, 
sub-populations of globular clusters. The dotted line represents the
median value for the $r_h$ of each sub-population.}
\end{figure}

%---------------------------------------------------------------------

\end{document}